\documentclass[a4paper,aps,pre,twocolumn,groupedaddress,showkeys,showpacs,
  floats,floatats,floatfix]{revtex4}
\usepackage[utf8]{inputenc}
\usepackage[english]{babel}
\usepackage{graphicx}
\usepackage{dcolumn} 
\usepackage{bm}
\usepackage{natbib}
\usepackage{latexsym}
\usepackage{mathrsfs}
\usepackage{amssymb}
\usepackage{amsmath}
\usepackage{amscd}
\usepackage{color}
\usepackage{pifont}
\usepackage{pstricks,pst-node,pst-text,pst-3d}
\usepackage{verbatim}
\usepackage{ulem}
\usepackage[T1]{fontenc}
\usepackage{hyperref}
\hypersetup{
  colorlinks   = true, 
  urlcolor     = black, 
  linkcolor    = red, 
  citecolor    = blue 
}
\newrgbcolor{Red}{1.0 0.0 1.0}
\begin{document}
\title{Steering multiattractors to overcome parameter inaccuracy and noise
effects}
\author{Rafael M.~da Silva$^{1}$}
\email{rmarques@fisica.ufpr.br}

\author{Nathan S.~Nicolau$^{2}$}
\email{tanicolau@fisica.if.uff.br}

\author{Cesar Manchein$^{2}$}
\email{cesar.manchein@udesc.br}
\author{Marcus W.~Beims$^{1}$}
\email{mbeims@fisica.ufpr.br}
\affiliation{$^1$Departamento de F\'\i sica, Universidade Federal do Paran\'a, 
81531-980 Curitiba, PR, Brazil}
\affiliation{$^2$Departamento de F\'\i sica, Universidade do Estado de Santa 
Catarina, 89219-710 Joinville, SC, Brazil} 
\date{\today}
%
\begin{abstract} Steering of attractors in multistable systems  is used to increase the 
available  parameter domains which lead to stable dynamics in nonlinear physical systems, 
reducing substantially undesirable effects of parametric inaccuracy and noise.
The procedure proposed here uses \,{time and/or space asymmetric} 
perturbations to move {\it independent} multistable attractors in phase space. Applying
this mechanism we increase around $230\%$ the stable domains in H\'enon's 
map, roughly $85\%$ 
in the ratchet current described by the Langevin equation and $60\%$ in Chua's 
electronic circuit. The proposal is expected to have wide applications 
in generic nonlinear complex system presenting multistability, so that related 
experiments can increase robustness under parametric inaccuracy and noise.
\end{abstract}
%
\pacs{05.45.Ac,05.45.Pq}
\keywords{Multiattractors, ratchet currents, transport, electronic devices.} 
\maketitle

{\it Introduction}.Any process in nature may be affected by intrinsic parameter
inaccuracy and noise, specially at smaller scales where such intrinsic 
effects can become the dominant processes. Inaccuracy in the values of 
parameters may destroy the desired dynamics of the system under analysis. For 
example, the resistance value in an electronic circuit is a parameter. Due to 
manufacture inaccuracy of the resistance value, {\it i.~e.~}the mean deviation 
of the normal produced resistance value, the electronic circuit can be 
affected leading to an undesirable dynamics \cite{Chaos-26-083107-2016}. 
Parameter inaccuracy is a remarkably general property existent in the 
estimation of cardiovascular parameters \cite{fan18}, in optical parameters in 
terahertz time-domain spectroscopy \cite{missori17}, in the charge carrier 
mobility in conjugated polymers \cite{ivo17}, in estimating parameters in 
biological networks \cite{craft17}, in the effective mass between  particles 
in classical \cite{her08} and quantum \cite{tutuc04} wells, among many others.

A key development in the description of nonlinear complex systems was the 
discovery of Stable Structures (SSs) in parameter space. When 
parameters are chosen inside such SSs, the underline dynamics is Lyapunov 
stable, periodic and determines the complex behavior. Thus, the 
SSs, observed in many systems
\cite{kapral82,markus89,mira91,jasonPRL93,broer98,bonatto05,kurths06,
bonattoR07,gallas-diode10,gallas-gallas-gallas,stoop10,rene11,gon13,denis11,
diego11,alan11-1,alan11-2,rene16,gallasCSF13},
delimit the range of available parameters which lead to stable dynamics.
Even though the behaviour found inside SSs is desirable in many situations, 
there exist two huge problems: (i) due to the parameter 
inaccuracy mentioned above, parameters may change and are not anymore 
necessarily located inside the boundaries of the SSs and (ii) the SSs are 
easily destroyed due to inevitable noise effects from the surrounding 
environment. In general, noise effects start to destroy the SSs 
from their borders \cite{carlo12,alan13-1,alan15,carlo16,ana17}. 
The relevant question then evolves into: is there any procedure which 
allows experiments {and simulations in complex systems} to run 
properly despite the parameter inaccuracy  and noise 
effects? This work proposes a mechanism to answer this positively. 
Instead trying to reduce the undesirable consequences of  parameter inaccuracy,
we propose to 
enlarge the parameter domains which generate the desired dynamics, {\it i.e.} 
to enlarge the SSs generating a flexibility in the allowed parameter 
values. Consequently the relative destructive effect  of noise is reduced
\,{increasing the critical noise intensity which destroys the SS. 
In this sense we increase the structural stability, if any}.

\begin{figure}[!b]
  \includegraphics*[width=1.0\columnwidth]{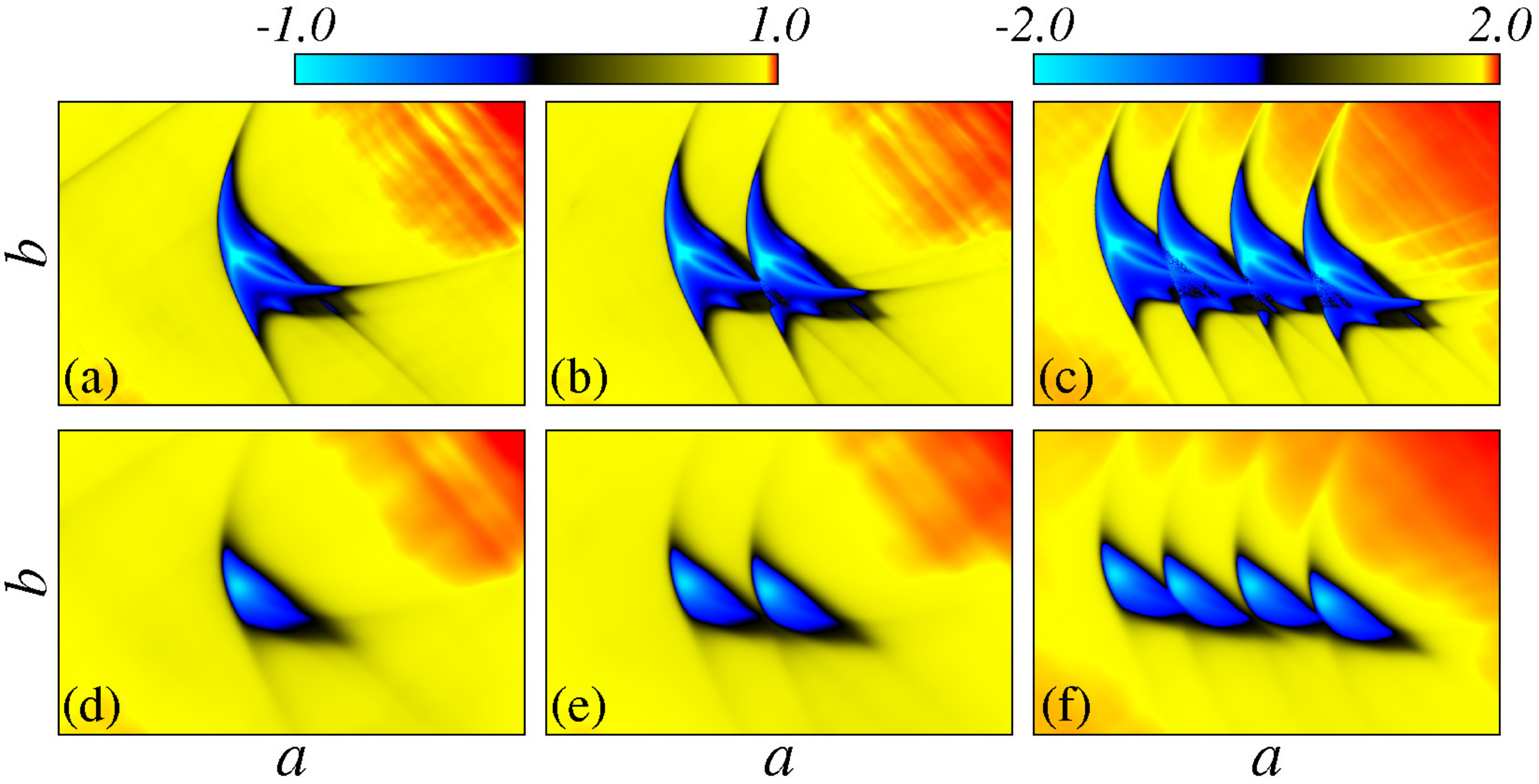}
  \caption{Enlargement of H\'enon's stable domains in two-dimensional
  parameter space ($a,b$) inside the interval $(a_{\mbox{\tiny min}},
  a_{\mbox{\tiny max}})=(1.713,1.743)$ and $(b_{\mbox{\tiny min}},
  b_{\mbox{\tiny max}})=(0.110,0.119)$ for noise intensities $7\times 10^{-5}$ 
  (top row) and $2\times 10^{-4}$ (bottom row).}
  \label{rat}
\end{figure}

Being more specific, to motivate and exemplify our proposal, Fig. \ref{rat} 
displays the enlargement of stable domains in the two-dimensional
parameter space ($a,b$) of the H\'enon map $(x_{n+1},y_{n+1})=(1-ax_n^2+y_n+F_n 
+ D\phi_n,bx_n)$ subjected to an external periodic force $F_n$ and a 
Gaussian noise $\phi_n$ with intensities proportional to
$D=7\times 10^{-5}$ (top row) and $D=2\times 10^{-4}$ (bottom row). 
Blue colours are related to parameters which induce stable dynamics and yellow 
to red colours leading to chaotic dynamics. Plotted is the largest Lyapunov 
exponent (LE) (see colour bar). The stable domains are the SSs mentioned before 
but slightly deformed by noise \cite{rafael17-2,ana17}. Going from 
Fig.~\ref{rat}(a) ($F=0$) to Fig.~\ref{rat}(b) ($F=8\times 10^{-4}$) 
and finally to Fig.~\ref{rat}(c) ($F=2\times 10^{-3}$), an astonishing 
increase of stable domains is observed. The increasing area of the stable domains 
are, when compared to Fig.~\ref{rat}(a), $85\%$ in Fig.~\ref{rat}(b) and 
$230\%$ in Fig.~\ref{rat}(c). The same behaviour is 
observed for the bottom row (same values of $F$) for which the gain is of $77\%$ 
for duplication [Fig.~\ref{rat}(e)] and $205\%$ for quadruplication 
[Fig.~\ref{rat}(f)] when compared to Fig.~\ref{rat}(d). The origin of such 
enlargement is that {\it independent attractors are steered}  to 
distinct locations in phase space. Two attractors in Figs.~\ref{rat}(b) and (e) 
and four in Figs.~\ref{rat}(c) and (f). In this case the attractors 
are created and steered by the time dependent function $F_n$ which has period-2 
($+F,-F,+F,-F,\ldots$) in Fig.~\ref{rat}(b) and (e) and period-4 ($+F,-F/2,+F/4,
-F,+F,-F/2,\ldots$) in Fig.~\ref{rat}(c) and (f). The physical and mathematical 
backgrounds of our findings are demonstrated next using realistic systems.
\,{It is worth to mention here that in the above motivational example of
the H\'enon model, the multiattractors attractors are {\it simultaneously created}
and {\it moved} in phase space. The H\'enon map without external perturbation
has only one attractor for the considered parameters.}

This Letter shows that by steering multiple independent attractors in phase 
space, overlapped SSs split apart generating enlarged stable domains in 
parameter space, leading to a substantial enhanced resistance under parameter 
inaccuracy and noise. The replication of periodic windows in differential equations 
was previously reported in \cite{medeiros10,rene11}, but a methodology to apply this 
procedure remained unclear. Our procedure introduces the concept 
of {\it enhanced parameter flexibility under steering of attractors}
and  should be applicable to all experiments and complex systems
models whose underline dynamics  presents multistability.
Results are presented for two complete distinct physical situations, namely for 
the Ratchet current described by the Langevin equation and for the Chua's 
circuit. While investigations about applications of ratchet effects are still very 
active 
\cite{ding18,heck17,bram17,brox17,erbas17,gall17,hans16,olson16,budkin16,koenig16,denisov16,bao16}, 
the Chua's electronic circuit \cite{chua92}  contains all ingredients of regular 
and chaotic motion and is also of actual interest 
\cite{tall18,leonov17,shep17,pen17}. 

\begin{figure}[!t]
  \centering
  \includegraphics*[width=1.0\columnwidth]{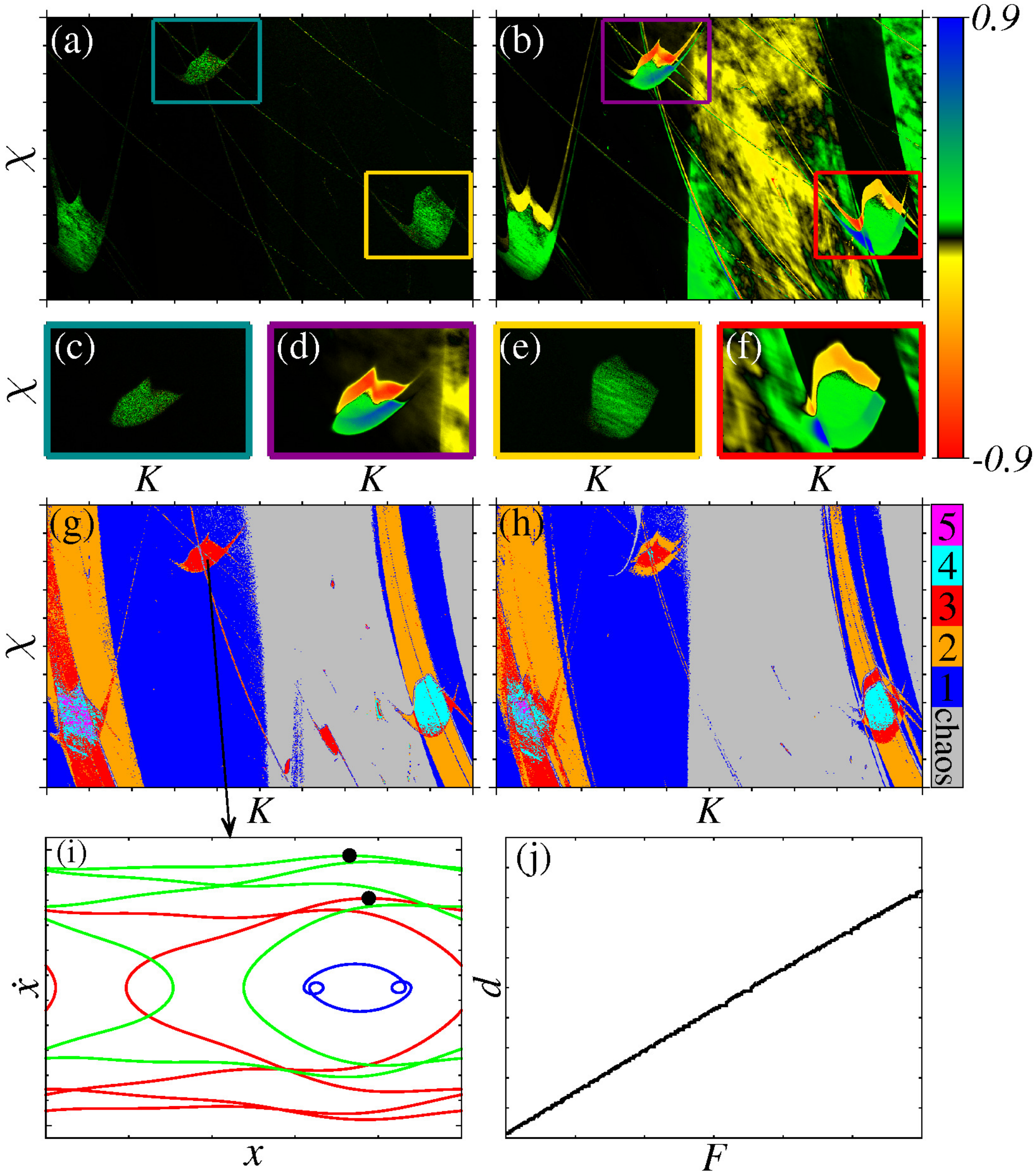}
  \caption{Two-dimensional parameter space $(K,\chi)$ for the 
  Langevin equation plotting the ratchet current in (a) $F=0.0$ and (b) $F=0.3$, 
  and the number of stable attractors in (g) for $F=0.0$ and (h) for $F=0.3$ 
  using $(K_{\mbox{\tiny min}},K_{\mbox{\tiny max}})=(5.6,9.6)$ and 
  $(\chi_{\mbox{\tiny min}},\chi_{\mbox{\tiny max}})=(0.7,0.9)$. Figures (c)-(f) 
  show thermal effects using $T=10^{-4}$ for the regions selected by coloured 
  boxes in (a) and (b). Figure (i) displays the stable attractors in phase-space 
  $(x_{\mbox{\tiny min}},x_{\mbox{\tiny max}})=(-\pi,\pi)$ and 
  $(\dot{x}_{\mbox{\tiny min}},\dot{x}_{\mbox{\tiny max}})=(-11,11)$, while (j) 
  shows the distance $d$ between the black dots in (i), which are the maxima of 
  velocity $\dot{x}$ for the attractors red and green, \,{in the ranges 
  $(F_{\mbox{\tiny min}},F_{\mbox{\tiny max}})=(0.0,0.3)$ and $(d_{\mbox{\tiny min}},
  d_{\mbox{\tiny max}})=(3.4,3.5)$}}
  \label{Nrrat}
\end{figure}

{\it Ratchet current in the Langevin equation.}
The most general way to describe unbiased currents in realistic systems
is to integrate the Langevin equation with an asymmetric spatial potential,
namely
%
%
$$\ddot{x}+\gamma\dot{x}-5.0[0.7\sin(2x)+\cos(x)]-K\sin(t)+\xi(t)=0,$$
%
where $x$ is the position, dot represents time derivative, $\xi(t)$ 
is the Gaussian thermal noise satisfying $\langle \xi(t)\rangle=0$ and 
obeying the dissipation-fluctuation relation 
$\langle \xi(t)^2\rangle = 2 \gamma K_{\mbox{\tiny B}} T \delta(t-t')$.
$K_{\mbox{\tiny B}}=1$ is the Boltzmann constant,
$\gamma$ is the dissipation parameter which induces time irreversibility, $T$ 
the temperature and $K\sin(t)$ is the force which keeps physical states out 
of equilibrium. For distinct parameters combination ($K,\chi=e^{-\gamma}$), 
used to construct two-dimensional parameter spaces studied here,
ratchet current may be generated, as shown in Fig.~\ref{Nrrat}(a) for $T=0$. 
To integrate the Langevin equation we use fourth order Stochastic
Runge-Kutta algorithm with fixed time-step $h=0.01$. Ratchet current is determined 
by double averages $\langle\langle\dot x\rangle\rangle$, one in time and the 
other one along $625$ equally distributed initial conditions (ICs) inside the 
interval $[-2\pi,2\pi]$, {\it i.e.} $\langle x(0) \rangle = \langle \dot x(0)
\rangle =0$. Colours represent the values of the current, black for zero, 
green to blue for positive and yellow to red for negative currents. It is evident 
that for almost all parameter combination no ratchet current is observed. 
Exceptions are three regions with small currents (green points). Two of these 
regions, marked with cyan and yellow boxes contain SSs,
not clearly visible due to the small current values.  These SSs are more 
complicated SSs than those usually studied in the literature (see introduction). 
The thermal noise tends to destroy a considerable portion of the SSs, 
always starting from their antennas, decreasing the parameter 
combinations that generate non-zero ratchet current. This is 
displayed in Figs.~\ref{Nrrat}(c) and (e) for $T=10^{-4}$.

In order to successfully apply our procedure we have to follow two fundamental 
steps. First is to check if the number of attractors for a parameter combination 
\,{of the unperturbed system}
is larger than one, {\it i.e.} if \,{there are} multiattractors 
\,{in the correspondent phase space}. This is displayed in Fig.~\ref{Nrrat}(g) 
for the same two-dimensional parameter space from Fig.~\ref{Nrrat}(a). In this 
case, colours are related to the number of periodic 
attractors, as shown in the colour bar. For each pair $(K,\chi)$ we 
calculated the time average $\langle \dot{x} \rangle$ for $25$ ICs equally 
distributed inside the interval $[-2\pi,2\pi]$ and, comparing them, it 
is possible to determine the number of different periodic attractors. Regions
with chaotic attractors are easily found analysing the largest LE. It is 
well known that usually stable attractors generate efficient
ratchet currents \cite{alan11-1,alan11-2}. Comparisons between Figs.~\ref{Nrrat}(g)
and (a) convince us that only regions with three and four stable attractors are 
able to generate the small currents. The blue and orange regions generate zero 
currents since attractors are located symmetrically around $\dot x=0$ (not 
shown). In order to explain the small currents in Fig.~\ref{Nrrat}(a) we focus 
on the three stable attractors located inside the cyan box. Figure 
\ref{Nrrat}(i) displays these attractors located at $K=7.1$ and $\chi=0.865$. It 
is easy to see that the red and green attractors are not located completely 
symmetrically around $\dot x=0$, thus generating the small currents inside the 
cyan box from Fig.~\ref{Nrrat}(a). 

Now we come to the next step, which is to steer the stable attractors to 
distinct directions in phase and parameter spaces. This can be achieved by 
inserting in the above Langevin equation an additional external force 
\,{
\begin{equation}
F(t)=F\cos(2t), 
\nonumber
\end{equation}}
which has a distinct period and symmetry from the external 
oscillation force $K\sin(t)$ which is already there. Other kind of periodic 
forms for $F(t)$ could be used, depending on specifics needs, as discussed later.
Figure \ref{Nrrat}(j) shows an example of how the distance $d$ between the maxima 
of $\dot{x}$ for two different attractors (see black dots in the attractors from 
Fig.~\ref{Nrrat}(i)) changes as a function of $F$. In Fig.~\ref{Nrrat}(h) the 
changes in the number of attractors in the two-dimensional parameter space 
can be observed for $F=0.3$. Attractors move in phase space and the SSs enlarge 
by a given amount. Not all of them move since this depends on the 
function $F(t)$, as mentioned above. Inside the periodic structures from the 
cyan box [compare Figs.~\ref{Nrrat}(a),(g) and (h)] some regions remain with 
only two attractors  from the three attractors shown in Fig.~\ref{Nrrat}(i). 
The effect of such movements on the current is quite interesting and is plotted 
in Fig.~\ref{Nrrat}(b) for $T=0.0$, and in Figs.~\ref{Nrrat}(d) and (f) for 
$T=10^{-4}$. To see the efficiency of the procedure proposed here we have to 
compare Fig.~\ref{Nrrat}(c) ($F=0$) with Fig.~\ref{Nrrat}(d) ($F=0.3$) and 
Fig.~\ref{Nrrat}(e) ($F=0$) with (f) ($F=0.3$). Figs.~\ref{Nrrat}(d) and (f) 
display enlarged SSs and much larger values of the currents when compared to 
the cases with $F=0$. The increase of the regions with non-zero ratchet currents 
is around $85\%$ when comparing Fig.~\ref{Nrrat}(a) and Fig.~\ref{Nrrat}(b). 
From the total number of attractors found for a specific parameter combination 
$(K,\chi)$, two of them, with opposite value of $\langle \dot{x}\rangle$ (time 
average of $\dot{x}$), are steered leading to two independent SSs, which have 
opposite currents (blue and red). Thus, the available currents domains in 
parameter space are enlarged by moving apart the attractors in 
Fig.~\ref{Nrrat}(i) with $\langle \dot{x} \rangle>0$ (green) and $\langle 
\dot{x} \rangle<0$ (red), avoiding the reduction of the total ratchet current 
when both attractors are found for same parameter combination $(K,\chi)$. 

{\it Chua's electronic circuit.} In this circuit the parameters are directly 
related to properties of the experimental device, namely, resistance, 
inductance and capacitances in the circuit. Due to experimental instabilities 
like electric voltage variations and noise, or even intrinsic imprecision in 
the resistance and capacitance values, there is no 
guarantee that the underline dynamics corresponds to the parameters for which 
the experiment was performed. According to the Kirchhoff's laws applied for 
current and voltage in the circuit, we find the following first order 
differential equations:
\renewcommand{\arraystretch}{0.3}
\begin{equation}
\begin{array}{cl} 
&\dot{X}=\dfrac{dX}{dT}=\alpha \left[Y-X-I(X)\right],\\
\\
&\dot{Y}=\dfrac{dY}{dT}=X-Y+Z,\\
\\
&\dot{Z}=\dfrac{dZ}{dT}=-\beta Y-\gamma Z-\left(\dfrac{\beta }{B_p}\right)F(X)
+ \phi(t).\\
\end{array}
\end{equation}
%
The piecewise linear current through the diode is $I(X) = b X + \frac{1}{2} 
(a-b) (\vert X + 1 \vert - \vert X -1 \vert)$, with $a$ and $b$ being control 
parameters and $F(X)=K[\sin(B_{p}X)+A\cos(3B_{p}X)]$,
which is the external asymmetric perturbation used to steer attractors. The  
adimensional states variables ($X,Y,Z$) are written in terms of the original 
states ($x,y,z$) as $X=x/B_p, Y=y/B_p$ and $Z=zR/B_p$ and parameters 
$\alpha=C_2/C_1$, $\beta ={R^2C_2}/{L}$, $\gamma ={Rr_LC_2}/{L}$, 
$T={t}/{RC_2}$, and $I(X)=i_d(x)R/B_p$.  The states ($x,y$) 
are the voltages across the capacitors $C_1$ and $C_2$ and $z$ the current 
across the inductor. $R$ and $L$ are passive linear elements and $r_L$ is the 
inductor-resistance. The fixed parameters are $a=-1.13996128$, $b=-0.7120006131$, 
$B_p=2.2$, $\beta=50.0$ and $K=0.01$, while $A$, $\alpha$ and $\gamma$ are 
varied. For $K=0$ the above system is the usual Chua's circuit \cite{chua92}.
For $A\ne 0$ the current asymmetry is introduced and leads to the steering of 
attractors. The effect of noise $\phi(t)$ is introduced in the $Z$ state 
variable and obeys a Gaussian distribution with $\langle \phi(t) \rangle=0$ and 
\,{$\langle \phi(t)^2\rangle = 2 D \delta(t-t')$, being $D$ its intensity}. Noise 
effects in the original Chua's circuit is by itself an actual research subject 
\cite{flavio18-1}.
%
\begin{figure}[!hb]
  \centering
  \includegraphics*[width=0.99\columnwidth]{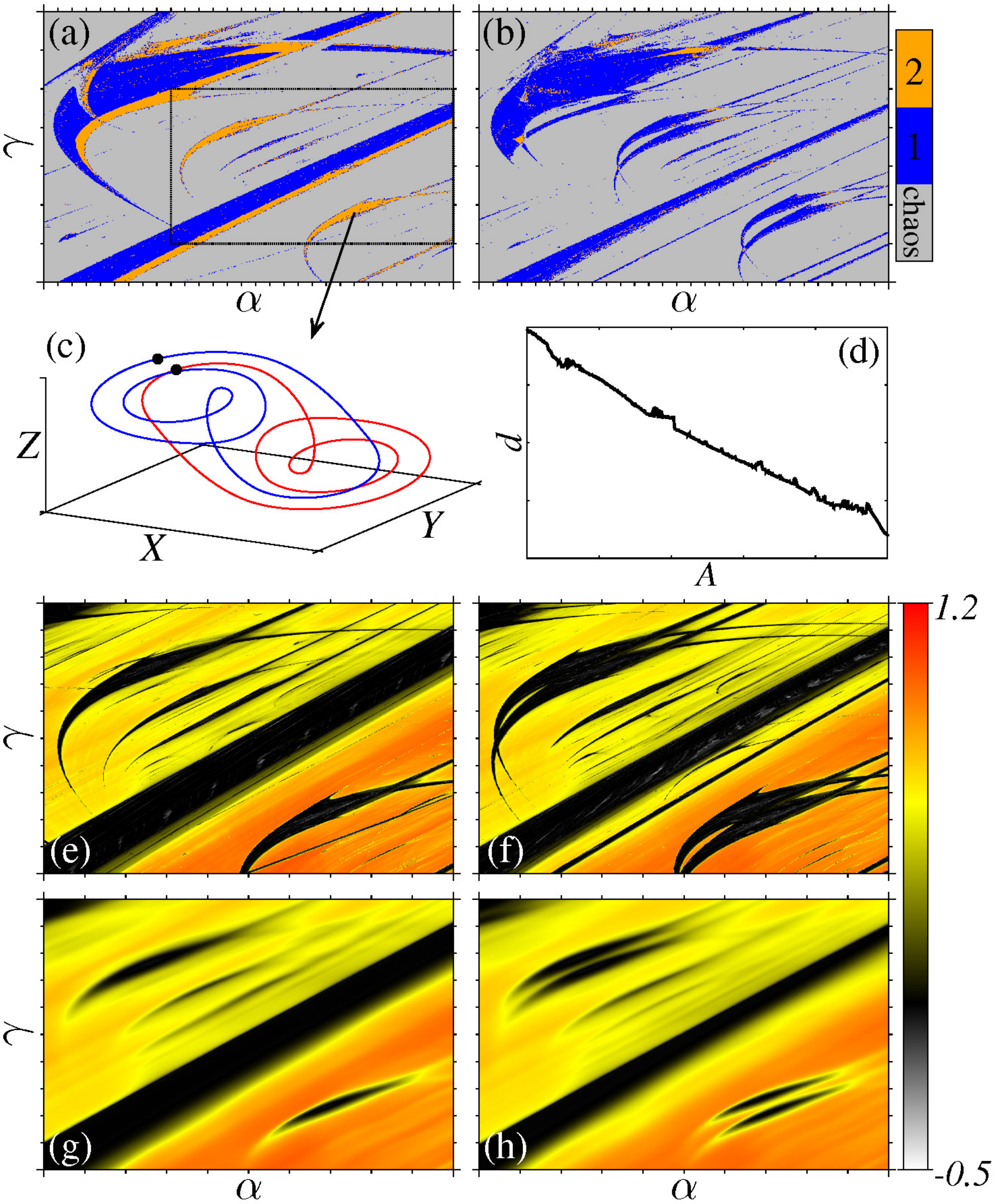}
  \caption{(Color online) Two-dimensional parameter spaces for Chua's 
  circuit showing the number of stable attractors in (a) $A=0.0$ and (b) $A=0.3$ 
  for $(\alpha_{\mbox{\tiny min}},\alpha_{\mbox{\tiny max}})=(21.8,24.8)$ and 
  $(\gamma_{\mbox{\tiny min}},\gamma_{\mbox{\tiny max}})=(-0.40,0.05)$. Panel 
  (c) displays the two attractors found in the SS indicated by the black 
  arrow {in the phase space limited by $(X_{\mbox{\tiny min}},
  X_{\mbox{\tiny max}})=(-2.5,2.5)$, $(Y_{\mbox{\tiny min}},
  Y_{\mbox{\tiny max}})=(-0.4,0.4)$ and $(Z_{\mbox{\tiny min}},
  Z_{\mbox{\tiny max}})=(-5.0,5.0)$, while (d)} shows the distance in coordinate 
  $z$ between the black points as function of $A$ \,{using $(A_{\mbox{\tiny min}}, 
  A_{\mbox{\tiny max}})=(0.0,0.15)$ and $(d_{\mbox{\tiny min}},d_{\mbox{\tiny max}})
  =(0.595,0.625)$}. Figures (e)-(h) are magnifications of the black box displayed 
  in (a) plotting the largest LE using (e) $A=0.0$, $D=0.0$, (f) $A=0.2$, $D=0.0$, 
  (g) $A=0.0$, $D=10^{-5}$ and (g) $A=0.2$, $D=10^{-5}$.}
  \label{chua}
\end{figure}

As before, the first step in the procedure is to find the parameter combination 
for which there are more than one independent stable attractors in phase space.
In this case, different stable attractors for a specific parameter combination 
$(\alpha,\gamma)$ are identified comparing the value of the lowest LE
(related to the most stable direction) obtained for $64$ different ICs equally distributed in 
$(x_{\mbox{\tiny min}},x_{\mbox{\tiny max}})=(y_{\mbox{\tiny min}},
y_{\mbox{\tiny max}})=(z_{\mbox{\tiny min}},z_{\mbox{\tiny max}})=(-0.2,0.2)$.
This is shown in Fig. \ref{chua}(a) in the two-dimensional parameter space 
($\alpha,\gamma$) for the noiseless case and $A=0$. Blue for one, orange for two 
stable attractors and gray for chaotic attractors. For this system only regions 
with one and two attractors were found inside this parameter range. We call to 
attention that the structures of the stable regions are the shrimp-like SSs 
frequently found in the literature. Figure \ref{chua}(c) shows exemplary two 
attractors in phase space for ($\alpha=24.06,\gamma=-0.31$) (see black arrow in 
Fig.\ref{chua}(a)). These attractors are obtained from distinct initial 
conditions and are therefore independent. Now we include the current asymmetry 
by using $A=0.3$ and the number of attractors for this case is displayed in 
Fig.\ref{chua}(b), demonstrating that only SSs with two attractors become 
separated. Each of the separated SSs have now only one stable attractor and the 
relative movement $d$ [distance between black points in Fig. \ref{chua}(c)] of 
attractors as a function of $A$ is displayed in Fig. \ref{chua}(d). 

The next step is to measure the enlargement of SSs for $A\ne0$ in the presence 
of noise. For clarity we analyse the enlargement inside the box shown in Fig. 
\ref{chua}(a). Figure \ref{chua}(e) for $A=0$ and Fig. \ref{chua}(f) for $A=0.2$ 
compare the enlargement in parameter space obtained by steering the stable 
attractors for the noiseless case, while Figs. \ref{chua}(g) and (h) compare the 
case with $D=10^{-5}$. Plotted in colours is the largest LE, black to white for 
stable motion and yellow to red for unstable motion. We observe that while the 
SSs with two attractors from Fig. \ref{chua}(a) become enlarged, the central 
black stripe which has one attractor starts to be destroyed. This show the 
relevance to having SSs with more than one attractor. The increased
area with stable dynamics in this case is around $60\%$ when comparing regions 
close to the duplicated SSs.

{\it Discussion.} The key procedure in the mechanism of steering 
multiattractors is to use \,{asymmetric time and/or space } external 
forces capable of moving them 
independently in phase space. \,{A simple systematic scheme 
to apply our method is: (i) run the simulation/experiment for the desirable 
parameter combination in a regular regime, (ii) plot the stable attractor in 
phase space (or coexisting stable attractors, in case), (iii) insert an 
asymmetric (in time or/and in space) periodic external perturbation and (iv) 
run the simulation/experiment again, plot the attractor and compare it to the 
attractor found in item (ii) above. Check if the attractor is split in more 
attractors (in case the original attractor was degenerated) or if the 
coexisting attractors moved apart \,{or got closer}. If yes, you 
achieved the needed extension of the available parameter which induce the 
same dynamics. If not, techniques to generated multiattractors in continuous 
systems \cite{PhysRevE.64.036223} might be 
used. In this case, start again in step (i). In the crucial step (iii) 
there is a maximum of $D+1$ possibilities, where $D$ is the 
spatial dimension of the system. The method is very general (universal) since 
there is no need to know the symmetry of the attractors}. 
\,{In addition, when the external force moves multiattractors equally in 
phase space, no enlargement of SSs is expected to occur. On the other hand,} 
when such external forcing is chosen with appropriate 
parameters (amplitude and frequency) it may be used to suppress SSs in parameter 
space \cite{rech12}, exactly the opposite of what is proposed here. 
We also have to mention a crucial difference between both continuous systems 
discussed here and the discrete H\' enon map used in the motivation. While in the 
continuous cases the extra external force only steers the {\it already existing} 
multistable attractors, in the map it simultaneously {\it creates} and 
{\it moves} them in phase and parameter space, as demonstrated for 
one-dimensional \cite{rafael17-1} and two-dimensional  \cite{rafael17-2} 
noiseless maps.

{\it Conclusions.} The steering of multistable attractors in phase space is
proposed to enlarge the available parameters of physical devices which lead to 
a desired dynamics. This could be interpreted as a kind of chaos control 
\cite{ott90-2} in parameter space, but having the fundamental principle 
of moving multiattractors apart \,{or closer to each other}. The enlarged stable 
domains (SSs) in parameter space, \,{only possible due to the existence of
{\it multiattractors}}, transform the underline dynamics more resistant to 
parameter inaccuracy and noise. Our procedure is motivated using the paradigmatic 
H\'enon map with noise, and explained in details for the ratchet currents in 
a thermal bath and Chua's electronic circuit with noise. The increasing 
percentage ($230\%$) of stable domains in H\'enon's map  and regions with 
non-zero current ($85\%$) in the ratchet system  are astonishing.  In Chua's 
electronic circuit the percentage gain is around $60\%$.  While results 
for the H\'enon map show the generality of our procedure in the context of 
discrete nonlinear systems, the ratchet current and Chua's circuit represent 
realistic systems described by continuous differential equations under noise. 
Thus, our mechanism should be applicable to a wide range of physical systems 
whose underline dynamics presents multistability. In case the systems have 
just one attractor, hidden attractors may be found \cite{leonov17,cubero16} or 
multiattractors \,{might} be created  \cite{PhysRevE.64.036223}. 
Our proposal is also expected to be useful for 
experiments in distinct areas, ranging from electronic circuits, lasers, ratchet 
devices, Josephson junctions, population evolutions, fluid dynamics, neuronal 
models, among others. Besides the experimental demonstration of our findings, 
future investigations may analise steering effects of independent chaotic 
multiattractors.

\acknowledgments{R.M.S.~thanks CAPES (Brazil) and C.M.~and 
M.W.B.~thank CNPq  (Brazil) for financial support. The  authors  also  
acknowledge computational support from Professor 
C.~M.~de Carvalho at LFTC-DFis-UFPR.}


\end{document}